\documentclass[prd,twocolumn,nopacs,floatfix,amsmath,nofootinbib,amssymb,preprintnumbers]{revtex4-2}
\usepackage{graphicx,subfigure,color,dcolumn,booktabs}
\usepackage{txfonts}
\usepackage{slashed}
\usepackage{epstopdf}
\usepackage{graphicx,xcolor,dcolumn,booktabs}
\usepackage[colorlinks,
            citecolor=blue,
            anchorcolor=red,
            menucolor=red,
            linkcolor=red,
            filecolor=red,
            runcolor=red,
            urlcolor=blue,
            frenchlinks=red]{hyperref}
\graphicspath{{Figures/}}

\begin{document}

\title{Likelihood analysis of the newly observed $f_0(2020)$, $f_0(2330)$, and $f_0(2470)$ in $J/\psi\to \gamma\eta^\prime\eta^\prime$ as high-lying unflavored scalar mesons}
\author{Cheng-Xi Liu$^{1,2,3,4}$}\email{liuchx2023@lzu.edu.cn}
\author{Li-Ming Wang$^{3,5}$}
\email{lmwang@ysu.edu.cn}
\author{Ting-Yan Li$^{1,2,3,4}$}\email{lity2023@lzu.edu.cn}
\author{Xiang Liu$^{1,2,3,4}$}\email{xiangliu@lzu.edu.cn}
\affiliation{
$^1$School of Physical Science and Technology, Lanzhou University, Lanzhou 730000, China\\
$^2$Research Center for Hadron and CSR Physics, Lanzhou University and Institute of Modern Physics of CAS, Lanzhou 730000, China\\
$^3$Lanzhou Center for Theoretical Physics, Key Laboratory of Theoretical Physics of Gansu Province and MoE Frontiers Science Center for Rare Isotopes, Lanzhou University, Lanzhou 730000, China\\
$^4$Key Laboratory of Quantum Theory and Applications of MoE, Lanzhou University,
Lanzhou 730000, China\\
$^5$Key Laboratory for Microstructural Material Physics of Hebei Province, School of Science, Yanshan University, Qinhuangdao 066004, China}

\begin{abstract}
Inspired by the newly observed three scalar states $f_0(2020)$, $f_0(2330)$, and $f_0(2470)$ by the BESIII Collaboration in $J/\psi\to \gamma\eta^\prime\eta^\prime$, we 
carried out the study of spectroscopic behavior of these high-lying unflavored scalar mesonic states. In this work, using the phenomenological description of radial $(n,M^2)$ trajectory and the quark pair creation model, we discussed the assignments of the $f_0(2020)$, $f_0(2330)$, and $f_0(2470)$ associated with the $f_0(2200)$ as the isoscalar high-lying scalar mesonic states and predicted the spectroscopic properties of their high-lying partners. The present study may provide valuable information for the construction of the scalar meson family. 
\end{abstract}

\maketitle

\section{Introduction}\label{sec1}

Light mesons have always been an important part of the whole hadron family, once inspiring the SU(3) classification of hadrons as a typical example~\cite{Gell-Mann:1964ewy}. Until now, the search for light mesons has been an ongoing task in hadron physics. In particular, with the accumulation of experimental data, more and more light flavor mesons with the masses around 2.2 GeV were found, including the $Y(2175)$ observed in $J/\psi\to \eta \phi f_0(980)$~\cite{BES:2007sqy} and the $X(2370)$ reported in the decay of $J/\psi\to \gamma \eta^\prime \pi^+\pi^-$~\cite{BESIII:2010gmv,BESIII:2019wkp}. These progresses show that there is a chance to construct the light meson family, which is also an effective approach to deepen our understanding of non-perturbative behavior of the strong interaction.

\begin{table}[htbp]
	\centering
	\setlength{\tabcolsep}{5mm}
	\caption{The resonance parameters of three $f_0$ states reported by BESIII.}
	\label{TAB1}
	\begin{tabular}{ccc}
		\toprule[0.5pt]
			\toprule[0.5pt]
		State & $m$ (MeV) & $\Gamma$ (MeV)\\
		\midrule[0.5pt]
		$f_0(2020)$ & $1982\pm 3^{+54}_{-0}$ & $436\pm 4^{+46}_{-49}$ \\
		$f_0(2330)$ & $2312\pm2^{+10}_{-0}$ & $134\pm 5^{+30}_{-9}$   \\
		$f_0(2470)$ & $2470\pm4^{+4}_{-6}$ & $75\pm 9^{+11}_{-8}$  \\
		\bottomrule[0.5pt]
		\bottomrule[0.5pt]
	\end{tabular}
\end{table}

\begin{figure}[htbp]
	\centering
	\begin{tabular}{c}
		\includegraphics[width=0.46\textwidth]{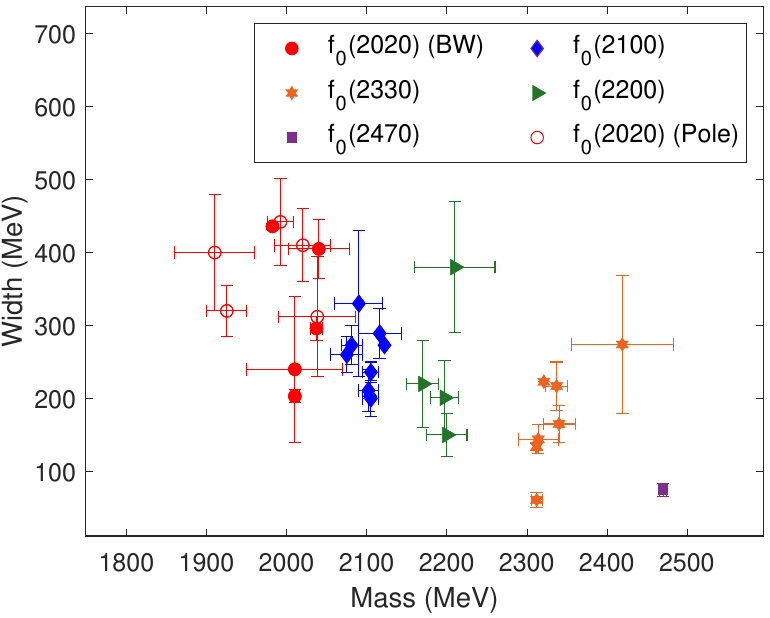}
	\end{tabular}
		\caption{
		A comparison of the resonance parameters of five isoscalar scalar states, {the $f_0(2020)$, including its pole mass  \cite{BESIII:2022zel,BESIII:2022iwi,Uman:2006xb,Anisovich:2000ut,GAMS:1998hws} and the Breit-Wigner mass
		\cite{Sarantsev:2021ein,Rodas:2021tyb,WA102:1999lqn,Ropertz:2018stk,WA102:1997sum}, }$f_0(2330)$ \cite{BESIII:2022iwi,BESIII:2022zel,Rodas:2021tyb,Sarantsev:2021ein,Bugg:2004rj,Anisovich:2000ut,Hasan:1994he}, $f_0(2470)$ \cite{BESIII:2022zel}, $f_0(2100)$ \cite{BaBar:2021fkz,BESIII:2013qqz,BES:1999dmf,Sarantsev:2021ein,Dobbs:2015dwa,Dobbs:2015dwa,Uman:2006xb,Anisovich:2000ut,Anisovich:1999ag,Bugg:1995jq,Hasan:1994he} and $f_0(2200)$ \cite{BES:2005iaq,DM2:1987ksj,Sarantsev:2021ein,Dobbs:2015dwa,Dobbs:2015dwa,Binon:2004yd,Hasan:1994he}. For these discussed scalar states, there exist different measurements of the resonance parameter for the same state, which are shown here. 
		{Here, the resonance parameters extracted from the pole positions and from the Breit-Wigner formula are presented in this figure, with circles and points, respectively, for comparison.}
		}
	\label{comparison}
	\end{figure}

Recently, the BESIII Collaboration performed a partial wave analysis of the process $J/\psi\to \gamma\eta^\prime\eta^\prime$~\cite{BESIII:2022zel} and found three scalar mesonic states $f_0(2020)$, $f_0(2330)$, and $f_0(2470)$ in the $\eta^\prime\eta^\prime$ invariant mass spectrum~\cite{BESIII:2022zel}. Their measured resonance parameters are collected into Table \ref{TAB1}. These observations make information about the scalar states with masses above 2 GeV become more abundant. 
For the $f_0(2020)$ and $f_0(2330)$, we note other previous experimental measurements of them via other production processes \cite{Uman:2006xb,Anisovich:2000ut,GAMS:1998hws,Sarantsev:2021ein,Rodas:2021tyb,WA102:1999lqn,Ropertz:2018stk,WA102:1997sum,Rodas:2021tyb,Sarantsev:2021ein,Bugg:2004rj,Anisovich:2000ut,Hasan:1994he}, which are summarized by presenting their resonance parameters in Fig.  \ref{comparison}. 
As collected in the Particle Data Group (PDG) \cite{Workman:2022ynf}, five scalar states above 2 GeV have been found so far, where three newly observed scalar states in the $\eta^\prime\eta^\prime$ channel were included.
Faced with this new situation, we propose that it is an appropriate time to establish high-lying unflavored scalar mesons in combination with the current observations of the $f_0(2020)$, $f_0(2330)$, and $f_0(2470)$ in the $J/\psi\to \gamma\eta^\prime\eta^\prime$ decay~\cite{BESIII:2022zel}. 
In Fig.  \ref{comparison}, we also make a comparison of their resonance parameters from different experiments\footnote{{In most cases, when the masses and total decay widths of scalar resonances are reported, it is often unclear whether these values correspond to pole parameters or their Breit-Wigner counterparts. It is in the case of $f_0(2020)$, but not in others.}}. We notice that the resonance parameters from different experiments are different for the same state. This mess situation should be clarified by further studies. 

In this work, we first discuss the possible assignment to the newly observed $f_0(2020)$, $f_0(2330)$, and $f_0(2470)$~\cite{BESIII:2022zel} and the scalar states $f_0(2200)$ collected in the PDG \cite{Workman:2022ynf}, using the phenomenological description of radial $(n,M^2)$ trajectory, which is an effective and successful method applied to 
the quantitative study of the mass spectrum of light mesons \cite{Chew:1962eu,Anisovich:2000kxa}. We can find two typical $(n,M^2)$ trajectories for isoscalar scalar mesons\footnote{In our previous work \cite{Guo:2022xqu}, we once studied some low-lying scalar mesons by the phenomenological description of radial $(n,M^2)$ trajectory, which is inspired by the observed $a_0(1817)$ \cite{BESIII:2022npc}. For these isovector $a_0$ states, the $a_0(980)$, $a_0(1450)$, and $a_0(1817)$ can form a $(n,M^2)$ trajectory. For these isoscalar $f_0$ states, we find another $(n,M^2)$ trajectory formed by the $f_0(980)$, $f_0(1500)$, $X(1812)$, and $f_0(2100)$. Along this line, the discussed $f_0(2330)$ is on the same $(n,M^2)$ trajectory that started from the $f_0(980)$. In this work, we do not discuss the $f_0(2100)$ since this scalar state has been studied in Ref. \cite{Guo:2022xqu}, which is a good candidate for the third radial excitation of the $f_0(980)$.}. Here, the $f_0(2330)$ is the fourth radial excitation of the $f_0(980)$, while the $f_0(2020)$, $f_0(2200)$, and $f_0(2470)$ are the second, third and fourth radial excitations of the $f_0(1370)$. In this radial $(n,M^2)$ trajectory starting from the $f_0(1370)$, there is the $f_0(1770)$ as the first radial excitation of the $f_0(1370)$. 

Although we can give a classification to these observed scalar states, we need to test it further by investigating their two-body Okubo-Zweig-Iizuka (OZI) allowed strong decays, which can be compared with the experimental data. To achieve this goal, we use the quark pair creation (QPC) model \cite{LeYaouanc:1972ae}
in the concrete calculation. In this way, the total and partial decay widths can be obtained, providing valuable information for deciphering the properties of these discussed scalar meson states. 

In addition to discussing these observed scalar states, we have also taken the opportunity to predict several high-lying scalar mesons, namely the $f_0(2570)$, $f_0(2660)$, and the isovector $a_0(2375)$ and $a_0(2610)$. 
The search for these predicted scalar states will be an interesting research topic, which is important for establish high-lying scalar mesonic states.

This paper is organized as follows. After the introduction, we illustrate how to categorize these scalar states above 2 GeV to the scalar meson family by the phenomenological description of radial $(n,M^2)$ trajectory, which is given in Sec. \ref{sec2}. And then, in Sec. \ref{sec3} we further study the corresponding two-body OZI-allowed strong decay behaviors of these discussed states, and make comparisons with the experimental data. In this way, the assignment to them can be tested. Of course, we also predict some scalar states missing in experiment that can be searched for in the near future.  The paper ends with a summary. 

\section{Mass spectrum analysis}\label{sec2}

To decode the properties of these discussed scalar states, we should perform mass spectrum analysis by fitting these states into the $(n,M^2)$ trajectories. This method has been extensively applied to classify light mesons \cite{Guo:2022xqu,Wang:2019qyy,Anisovich:2000kxa,Guo:2019wpx}. Additionally, we can use the $(J,M^2)$ trajectories, also known as the Regge trajectories \cite{Chew:1962eu}, for further classification. The phenomenological description of radial $(n,M^2)$ trajectory shows that the masses of hadrons follow a pattern that can be written as \cite{Anisovich:2000kxa}
\begin{eqnarray} \label{nm}
	M^2=M_0^2+(n-1)\mu^2,
\end{eqnarray}
where $M_0$, $n$ and $\mu^2$ denote the ground state mass, radial quantum number and the slope parameter of the
trajectory, respectively. 

\begin{figure*}[htbp]
	\centering
	\begin{tabular}{c}
		\includegraphics[width=1\textwidth]{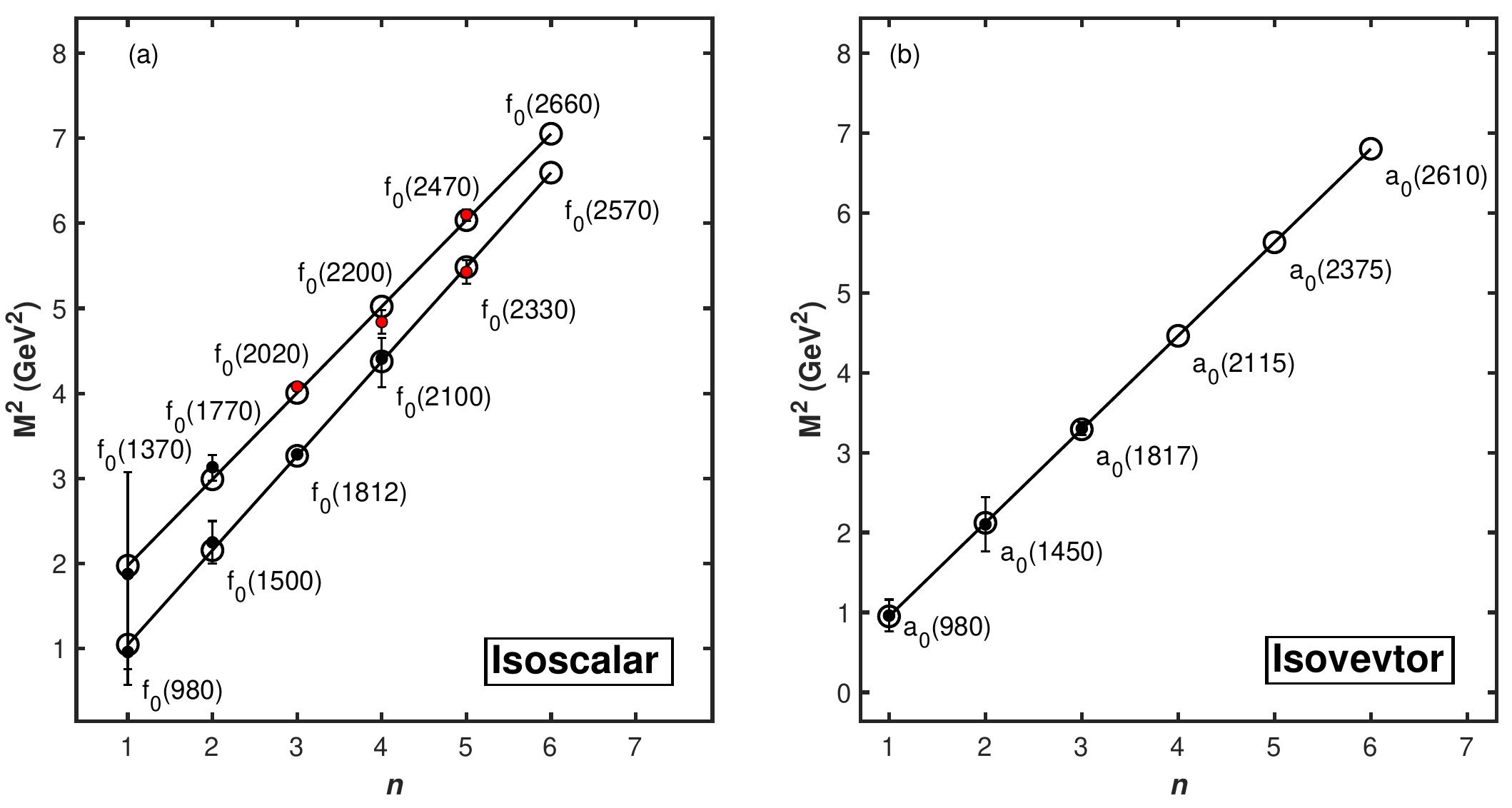}
	\end{tabular}
	\caption{
		Three $(n,M^2)$ trajectories of the scalar mesons family, including isoscalar $f_0$ states (a) and isovector $a_0$ states (b). Several scalar states discussed in this work are marked by red and black points and black circles. The red points mark states we study in this work, and the black circle mark states affirmed by the PDG \cite{Workman:2022ynf}. The black circles are the theoretical value
        of the $(n,M^2)$ trajectories and several states we predict. The data are taken from the PDG \cite{Workman:2022ynf} and BESIII result in Ref.~\cite{BESIII:2022zel}.  
	}
	\label{ReggePlot222}
\end{figure*}

As shown in the PDG, there are numerous scalar states capable of forming two $(n,M^2)$ trajectories for isoscalar scalar $f_0$ states and one $(n,M^2)$ trajectory for isovector scalar $a_0$ states (see Fig. \ref{ReggePlot222}). In Ref. \cite{Guo:2022xqu}, a $(n,M^2)$ trajectory was identified starting at the $f_0(980)$ and ending at the $f_0(2100)$ with a slope parameter of $\mu^2=1.13$ GeV$^2$. The presence of the $f_0(2330)$ is consistent with this $(n,M^2)$ trajectory, suggesting that it serves as the fourth radial excitation of the $f_0(980)$. Furthermore, we can anticipate the existence of the fifth radial excitation of the $f_0(980)$, denoted as the $f_0(2570)$ in this study.

Another $(n,M^2)$ trajectory with a slope parameter $\mu ^2=1.04$ GeV$^2$ can be formed from the $f_0(1370)$ by including the $f_0(1770)$, $f_0(2020)$, and the $f_0(2470)$ states, from which we can predict the $f_0(2660)$. Thus, the focused $f_0(2020)$ and $f_0(2470)$ are the good candidates for the second and fourth radial excitations of the $f_0(1370)$. 


In this work, we also present the $(n,M^2)$ trajectory with $\mu^2=1.17$ GeV$^2$ for isovector scalar states, depicted in Fig. \ref{ReggePlot222} (b). These states include the $a_0(980)$, $a_0(1450)$, and $a_0(1817)$, which are associated with the absence of the $a_0(2115)$, $a_0(2375)$, and $a_0(2610)$ in the trajectory\footnote{In Ref. \cite{Guo:2022xqu}, the properties of the $a_0(980)$, $a_0(1450)$, $a_0(1817)$, and the predicted $a_0(2115)$ were discussed, so here our focus shifts to the $a_0(2375)$ and $a_0(2610)$.}. Since the properties of the projected high-lying states $a_0(2375)$ and $a_0(2610)$ are still unknown, we will provide their two-body OZI-allowed strong decay information. It's noteworthy that the $f_0(2330)$, $f_0(2470)$, and the predicted $a_0(2375)$ belong to the same nonet. In addition, several predicted high-lying scalar mesonic states, namely the $f_0(2570)$, $f_0(2660)$, and $a_0(2610)$, are in the same nonet.

Before entering the formal discussion of their strong decay behavior, we summarize the possible assignments to these discussed states by the analysis of the $(n,M^2)$ trajectory (see Table \ref{nn}).  
In the following section, we are dedicated to the study of the two-body OZI-allowed strong decays of these discussed high-lying scalar states.

\begin{table}[htbp]
	\centering\label{nn}
	\setlength{\tabcolsep}{5mm}
	\caption{Summary of the possible assignments to these discussed scalar states given by mass spectrum analysis. Here, four predicted states are marked by underlines. $n$ denotes the radial quantum number corresponding to these scalar mesonic states.}
	\label{TAB2}
	\begin{tabular}{cccc}
		\toprule[0.5pt]
		$n=3$ & $n=4$ & $n=5$ & $n=6$ \\
		\midrule[0.5pt]
		&&$f_0(2330)$&$\underline{f_0(2570)}$\\
    	$f_0(2020)$&$f_0(2200)$&$f_0(2470)$&$\underline{ f_0(2660)}$ \\
    	&&$\underline{a_0(2375)}$&$\underline{a_0(2610)}$\\
		\bottomrule[0.5pt]
	\end{tabular}
\end{table}

\section{Two-body OZI-allowed strong decays}\label{sec3}

To obtain the two-body OZI-allowed strong decay information of these discussed high-lying scalar mesonic states, the quark pair creation (QPC) model \cite{LeYaouanc:1972ae,Blundell:1996as} utilized in this work  {has been successfully applied to the quantitative description of the OZI-allowed strong decay of hadrons \cite{Wang:2019qyy,Chen:2019ywy,Luo:2023sne,Pang:2017dlw,Wang:2016krl,Li:2020xzs,li:2021hss,Ni:2021pce,Li:2023wgq,Li:2022bre}.} 
A decay process $A\to B+C$ can be expressed as
\begin{eqnarray}
\langle BC|\mathcal{T}|A \rangle = \delta ^3(\mathbf{P}_B+\mathbf{P}_C)\mathcal{M}^{{M}_{J_{A}}M_{J_{B}}M_{J_{C}}},
\end{eqnarray}
where $\mathbf{P}_{B(C)}$ is a three-momentum of a meson $B(C)$ in the rest frame of a meson $A$. A superscript $M_{J_{i}}\, (i=A,B,C)$ denotes an orbital
magnetic momentum. The transition operator $\mathcal{T}$ is introduced to describe the creation of a quark-antiquark pair from the vacuum, which has the quantum number
$J^{PC}=0^{++}$, i.e., $\mathcal{T}$ can be written as
\begin{eqnarray}
\mathcal{T}& = &-3\gamma \sum_{m}\langle 1m;1~-m|00\rangle\int d \mathbf{p}_3d\mathbf{p}_4\delta ^3 (\mathbf{p}_3+\mathbf{p}_4) \nonumber \\
 && \times \mathcal{Y}_{1m}\left(\frac{\textbf{p}_3-\mathbf{p}_4}{2}\right)\chi _{1,-m}^{34}\phi _{0}^{34}
\left(\omega_{0}^{34}\right)_{ij}b_{3i}^{\dag}(\mathbf{p}_3)d_{4j}^{\dag}(\mathbf{p}_4).
\end{eqnarray}
This is fully constructed in the form of a visual representation to reflect the creation of a quark-antiquark pair from vacuum, where the quark and antiquark are denoted by the indices $3$ and $4$, respectively.
The parameter $\gamma$ represents the strength of the creation of $q\bar{q}$ from the vacuum. In this work we use $\gamma = 7.1$ in accordance with Refs. \cite{Guo:2022xqu,LeYaouanc:1977gm}.  $\mathcal{Y}_{\ell m}(\mathbf{p})={|\mathbf{p}|^{\ell}}Y_{\ell
m}(\mathbf{p})$ is the solid harmonics. $\chi$, $\phi$, and $\omega$ denote the spin, flavor, and color wave functions, respectively, which can be treated separately.
Subindices $i$ and $j$ stand for the color of a $q\bar{q}$ pair.

The decay amplitude can be expressed as another form using the Jacob-Wick formula \cite{Jacob:1959at}
\begin{eqnarray}
\mathcal{M}^{JL}(\mathbf{P})&=&\frac{\sqrt{4\pi(2L+1)}}{2J_A+1}\sum_{M_{J_B}M_{J_C}}\langle L0;JM_{J_A}|J_AM_{J_A}\rangle \nonumber \\
&&\times \langle J_BM_{J_B};J_CM_{J_C}|{J_A}M_{J_A}\rangle \mathcal{M}^{M_{J_{A}}M_{J_B}M_{J_C}},
\end{eqnarray}
then the general decay width reads as
\begin{eqnarray}
\Gamma&=&\frac{\pi}{4} \frac{|\mathbf{P}|}{m_A^2}\sum_{J,L}|\mathcal{M}^{JL}(\mathbf{P})|^2,
\end{eqnarray}
where $m_{A}$ is the mass of an initial state $A$. More details on the QPC model can be found in Refs. \cite{Pang:2014laa,vanBeveren:1979bd,Blundell:1996as,Li:2022khh,Ye:2012gu}.
In the calculation, the simple harmonic oscillator (SHO) wave function is adopted to depict the spatial wave function of these discussed scalar mesonic states.
{In momentum space, the SHO wave function is expressed as
\begin{equation}
\phi_{nlm}^p(R,{\bf p}) = R_{nl}^p(R,p)Y_{lm}({\bf\hat{p}}),
\end{equation}
where the radial component is given by
\begin{equation}
\begin{split}
R^p_{nl}(R,p)=&(-1)^n(-{\mathrm i})^l R^{\frac{3}{2}+l}\sqrt{\frac{2n!}{\Gamma(n+l+\frac{3}{2})}}L_{n}^{l+\frac{1}{2}}(R^2p^2){\mathrm e}^{-\frac{R^2 p^2}{2}}p^l .
\end{split}
\end{equation}
Correspondingly, in coordinate space, the wave function takes the form 
\begin{equation}
\phi_{nlm}^r(R,{\bf r}) = R_{nl}^r(R,r)Y_{lm}({\bf\hat{r}})
\end{equation}
with the radial component
\begin{equation}
\begin{split}
R^r_{nl}(R,r)=&\frac{1}{R^{\frac{3}{2}+l}}\sqrt{\frac{2n!}{\Gamma(n+l+\frac{3}{2})}}L_{n}^{l+\frac{1}{2}}({r^2}/{R^2}){\mathrm e}^{-\frac{r^2}{2R^2}}r^l.
\end{split}
\end{equation}
In these equations, $R$ is a parameter that scales the SHO wave function, and it can be determined by solving the  equation~\cite{Barnes:1996ff,Close:2005se}, i.e.,
\begin{equation}
\int |\phi_{nlm}^r(R,{\bf r})|^2{\bf r}^2{\rm d}^3{\bf r}=\int |\Phi({\bf r})|^2{\bf r}^2{\rm d}^3{\bf r},
\end{equation}
where the $\Phi({\bf r})$ represents the precise wave function obtained from the potential model~\cite{Close:2005se}. This approach is applicable to the well-established low-lying states. For the higher excited states examined in our study, we account for the $R$ dependence of the partial widths. Since the wave functions of the mesons involve the parameter $R$, the decay amplitude also dependents on $R$.}

{The model dependence of the QPC model primarily arises from the hadron wave functions and the 
$\gamma$ parameter. As discussed earlier, calculating the decay amplitude requires hadron wave functions as input, which are obtained by solving potential models. Thus, the model dependence of the QPC approach is largely rooted in the choice of potential models used to derive the wave functions. Additionally, the $\gamma$ parameter affects the decay widths, but in a straightforward manner—the decay widths are proportional to the square of $\gamma$. In previous studies, 
$\gamma$ values were determined by fitting known decay widths~\cite{Guo:2019wpx, Ye:2012gu, Pang:2017dlw}, and these values were subsequently used for making predictions.
}

\begin{figure*}[htbp]
	\centering
	\begin{tabular}{cc}
		\includegraphics[width=0.9\textwidth]{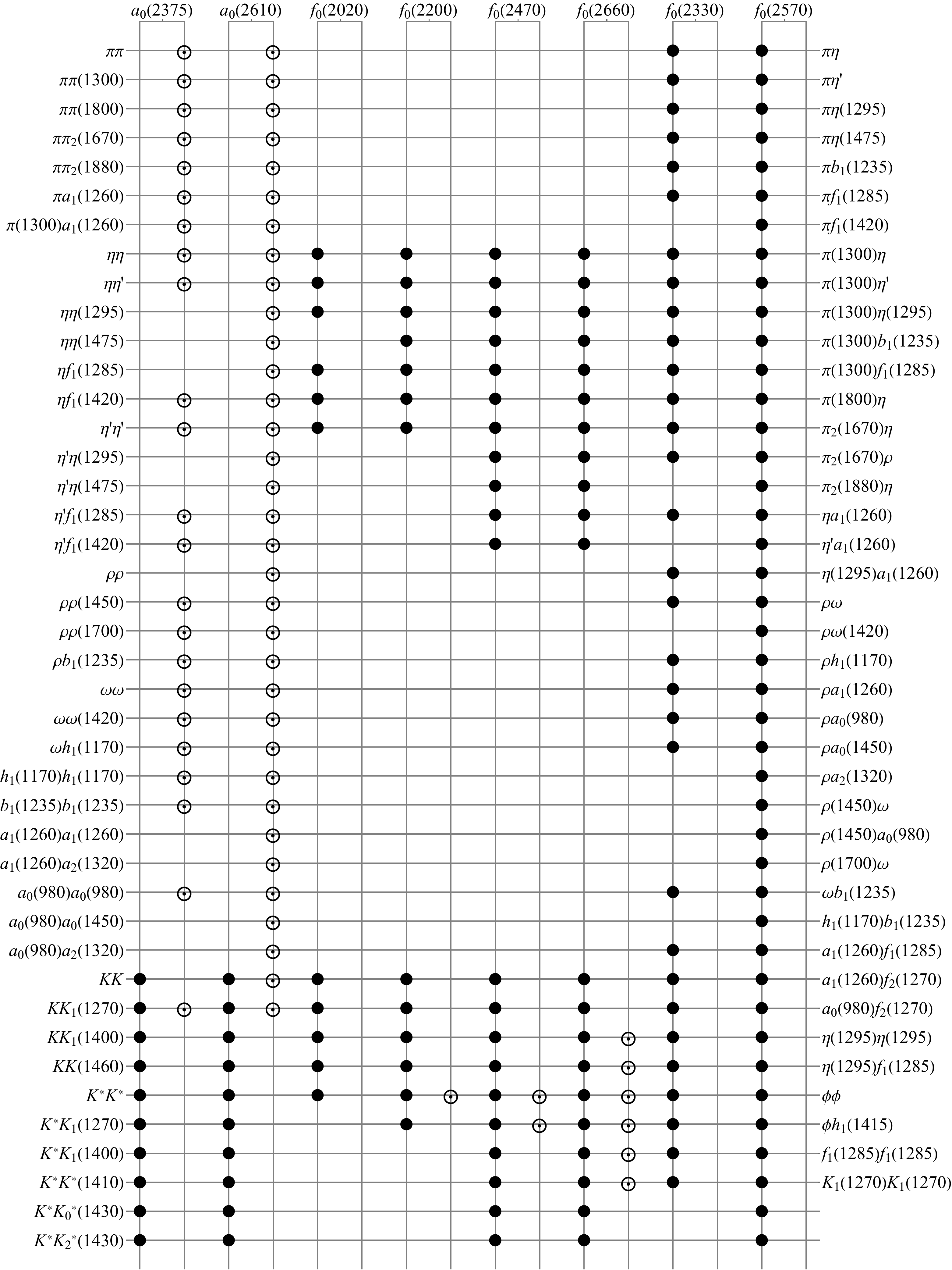}
	\end{tabular}
	\caption{
		The possible decay channels of these discussed scalar mesons above 2 GeV. Here, the black points denote the decay channels illustrated on the left, while the circle points 
		represent those shown on the right. {Here, we assume that these discussed states are either pure $n\bar{n}$ states or pure $s\bar{s}$ states. } 
	}
	\label{decay}
\end{figure*}

As shown in Fig.  \ref{decay}, we list these possible decay channels of the focused high-lying scalar mesons. Here, {it is assumed that the states in the trajectories initiated from the $a_0(980)$ and $f_0(980)$ are dominated by $n\bar{n}$ ($n\bar{n}$ denoting $(u\bar{u}+d\bar{d})/\sqrt{2}$), while those initiated from the $f_0(1370)$ are dominated by $s\bar{s}$. In the QPC model, for the $n\bar{n}$ states discussed, they can decay into a strange and an antistrange meson pair if $s\bar{s}$ quark pair creation from vacuum is considered. Similarly, the $s\bar{s}$ state can have these decay modes when $n\bar{n}$ is created from vacuum.}
In the following subsections, we present the numerical result of their strong decays and the corresponding discussions by comparing our results with the experimental data.

\subsection{The $f_0(2330)$ and the predicted $f_0(2570)$}


Radiative decays of $J/\psi$  provide a good platform for the study of light mesons \cite{Sarantsev:2021ein,Rodas:2021tyb}. A recent example is the observation of the $f_0(2330)$ in $J/\psi\to \gamma\eta^\prime \eta^\prime$ \cite{BESIII:2022zel},  which was first reported in the $p\bar{p}\to \pi\pi$ process with a mass of $M=2321$ MeV and a width of $\Gamma=223$ MeV \cite{Hasan:1994he}. This observation was further confirmed in the $p\bar{p}\to \pi\pi, \eta\eta$ channels \cite{Anisovich:2000ut}. The resonance parameters of the $f_0(2330)$ measured in different experiments, especially its width, show remarkable  discrepancies, as shown in Fig. \ref{comparison}.

\begin{figure*}[htbp]
	\centering
	\begin{tabular}{c}
		\includegraphics[width=1.0\textwidth]{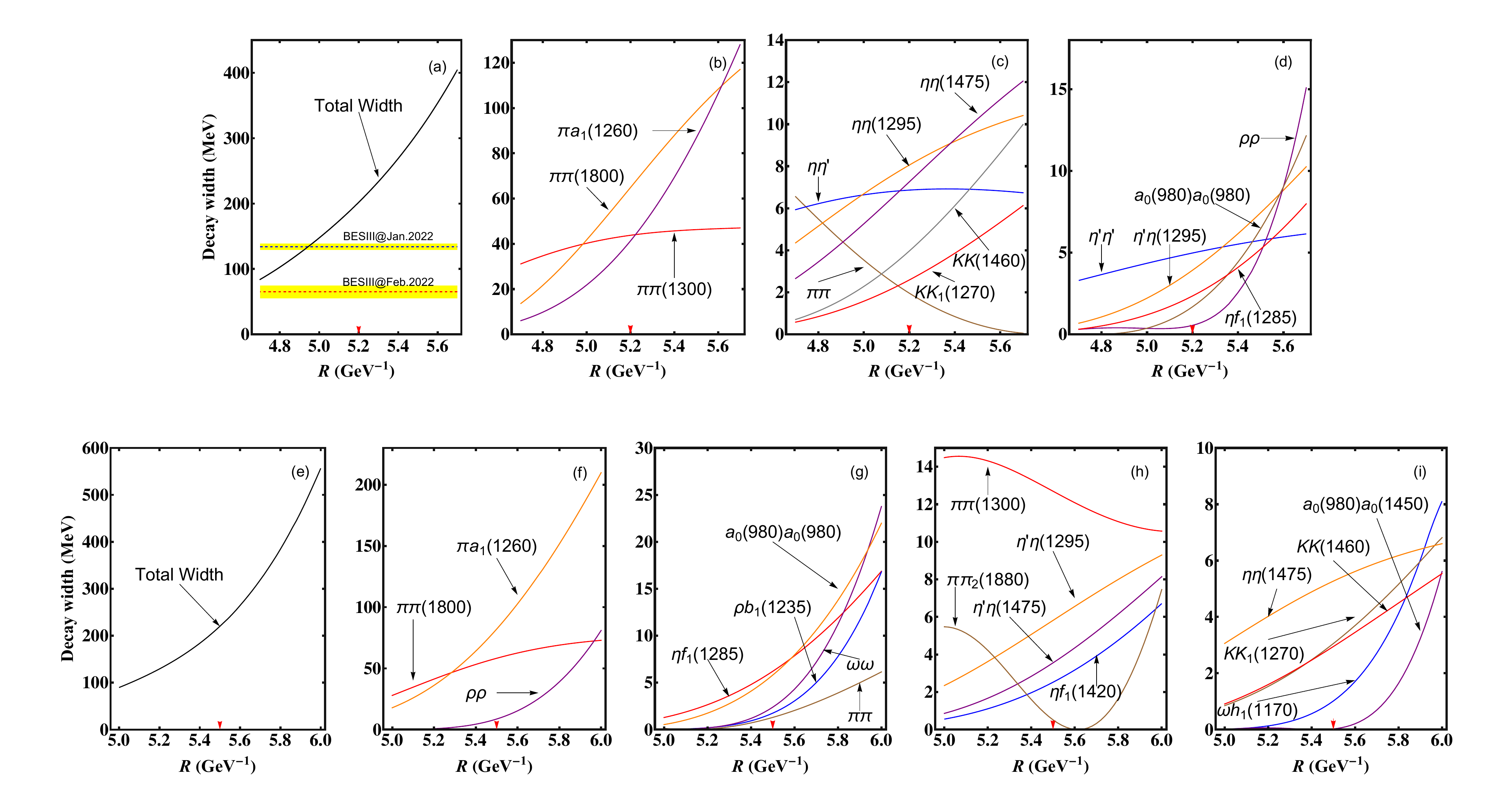}\\
	\end{tabular}
	\caption{(Color online.)
		The calculated partial and total decay widths of the $f_0(2330)$ (a)-(d) and the predicted $f_0(2570)$ (e)-(i) dependent on $R$ values. We compare our results with the measured width of the $f_0(2330)$ from BESIII \cite{BESIII:2022iwi,BESIII:2022zel} (see digram (a)). The channels with widths less than 5 MeV are not shown here, but their contributions are included in the total widths. Here, the corresponding calculated $R$ values for these discussed scalar mesons are also marked with red arrows.  }
	\label{f02330}
\end{figure*}

The $(n,M^2)$ trajectory analysis suggests that the $f_0(2330)$ is the fourth radial excitation of the $f_0(980)$, which can be tested by further investigating its strong decay allowed by the OZI rule. At present, experimental information on the $f_0(2330)$ decays remains scarce. In Fig. \ref{f02330}, we present the $R$ dependence of the calculated total and partial decay widths of the $f_0(2330)$. It is evident that the $\pi \pi(1300)$, $\pi \pi(1800)$, and $\pi a_1(1260)$ channels make the major contributions to the total decay width of the $f_0(2330)$. In particular, the partial decay widths involved in the $\pi \pi(1300)$, $\eta\eta^{\prime}$, and $\eta^{\prime}\eta^{\prime}$ channels are not sensitive to the $R$ value. Additionally, the $\eta^{\prime}\eta(1295)$, $\eta f_1(1285)$, $K K_1(1270)$, and other decay modes are its subordinate decay channels.



In Fig. \ref{f02330}, we also mark the typical $R$ values by the red arrows above the abscissa, which are calculated for each state as in Refs. \cite{Barnes:1996ff,Close:2005se}. Looking at the diagram in more details reveals a fluctuation in the width of the $f_0(2330)$, which corresponds to the variations in the $R$ value. However, the latest data given by BESIII \cite{BESIII:2022zel} are in agreement with the theoretical predictions. Our result can explain why the $f_0(2330)\to \eta^{\prime}\eta^{\prime}$ was observed by BESIII \cite{BESIII:2022zel}, since the $\eta^\prime \eta^\prime$ channel is sizable (see Fig. Fig. \ref{f02330} (d)).
Based on the above analysis, we may conclude that the $f_0(2330)$ is a good candidate for the fourth radial excitation of the $f_0(980)$.


In the following analysis, we also obtain the strong decay behavior of the $f_0(2570)$. The calculated total and partial decay widths are shown in Fig. \ref{f02330}. The total width of the $f_0(2570)$ ranges from 100 MeV to 500 MeV with variations in the proposed $R$ value ranging from $5.0-6.0$ GeV$^{-1}$. 
Here, the total width exhibits obvious fluctuations in response to variations in the $R$ value, mainly due to the node effect \cite{Duan:2020tsx}, which is particularly evident for higher radial excitations. 
In this work, we predict that the total width of the $f_0(2570)$ is about 200 MeV corresponding to the typical $R$ value of 5.5 GeV$^{-1}$, as denoted by the red arrow. Among the two-body strong decay channels, the $\pi\pi(1800)$, $\pi a_1(1260)$, and $\rho \rho$ channels contribute significantly to the total width. 
We suggest that future experiments focus on measuring the partial decay width of the $f_0(2570)$, which can serve as a test of the assignment of the $f_0(2570)$ as the fifth radial excitation of the $f_0(980)$.

\subsection{The $f_0(2020)$, $f_0(2200)$, $f_0(2470)$, and the predicted $f_0(2660)$}


The $(n,M^2)$ trajectory analysis suggests that the $f_0(1770)$ represents the first radial excitation of the $f_0(1370)$, which is considered to be the ground state. {The $f_0(2020)$ may occupy the position of the second radial excited state.} The $f_0(2200)$ and $f_0(2470)$ are identified as the third and fourth radial excitations within this family, respectively. Furthermore, we expect the existence of a missing state, the $f_0(2660)$, as the fifth radial excitation.


\begin{figure*}[htbp]
	\centering
	\begin{tabular}{c}
		\includegraphics[width=0.8\textwidth]{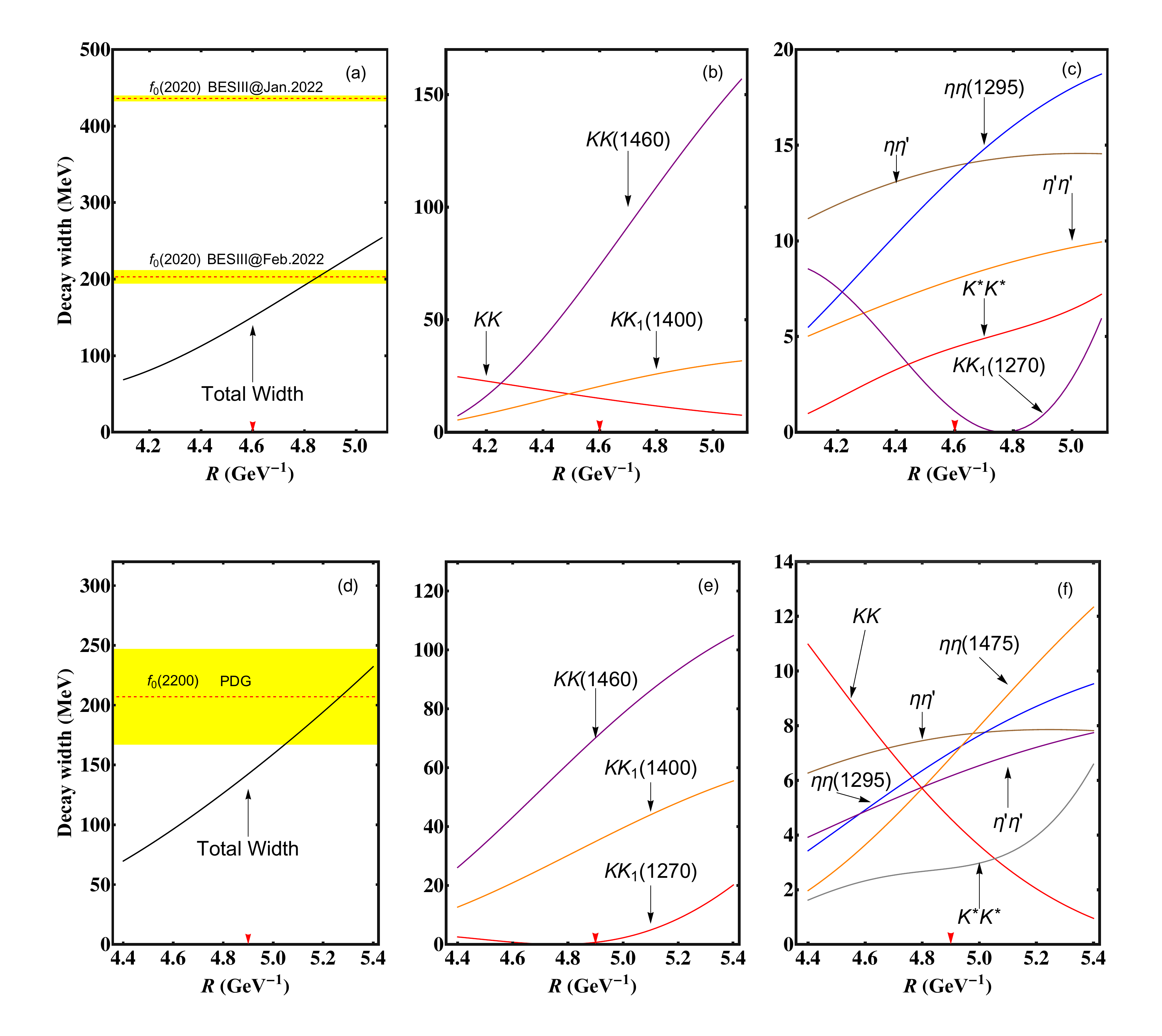}\\
	\end{tabular}
	\caption{(Color online.)
	{The calculated partial and total decay widths of the $f_0(2020)$ as a pure $s\bar{s}$ state (a)-(c) and the predicted $f_0(2200)$  (d)-(f) are dependent on the $R$ values. We also list the measured widths for the $f_0(2020)$ \cite{BESIII:2022zel,BESIII:2022iwi} and $f_0(2200)$ \cite{Workman:2022ynf} for the comparison with our results (see diagrams (a) and (d)). }The channels with widths less than 5 MeV are not shown here, but their contributions are included in the total widths. Here, the corresponding calculated $R$ values for these discussed scalar mesons are also marked with red arrows.}
	\label{f02020}
\end{figure*}

\begin{figure*}[htbp]
	\centering
	\begin{tabular}{c}
		\includegraphics[width=1.0\textwidth]{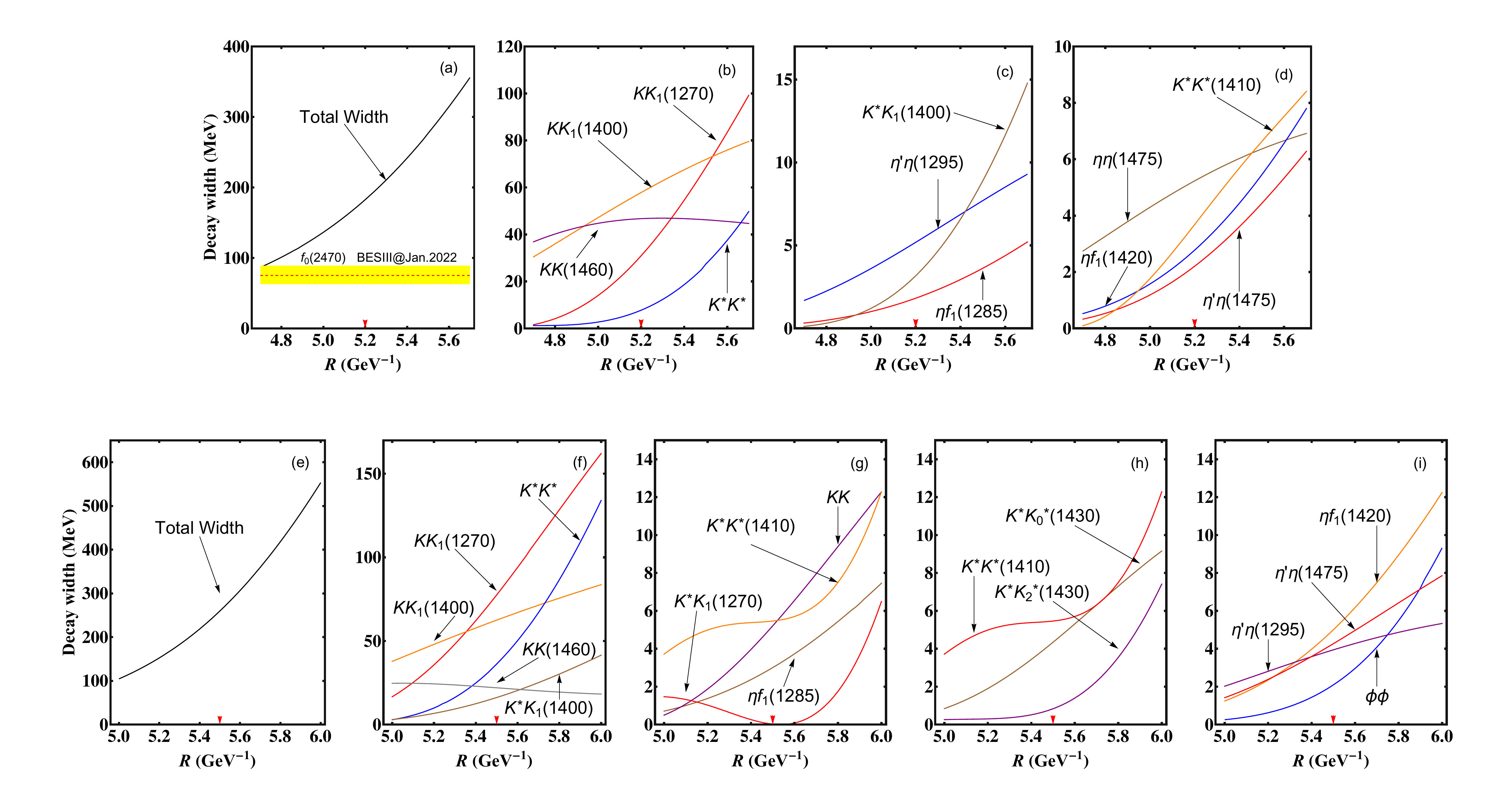}
	\end{tabular}
	\caption{(Color online.)
		The calculated partial and total decay widths of the $f_0(2470)$ (a)-(d) and the $f_0(2660)$ (e)-(i) dependent on $R$ values. We compare our results with the measured width of the $f_0(2470)$ from BESIII \cite{BESIII:2022zel} in diagram (a). The channels with widths less than 5 MeV are not shown here, but their contributions are included in the total widths. Here, the corresponding calculated $R$ values for these discussed scalar mesons are also marked with red arrows.}
	\label{f02470}
\end{figure*}

\begin{figure}[htbp]
	\centering
	\begin{tabular}{c}
		\includegraphics[width=0.46
	\textwidth]{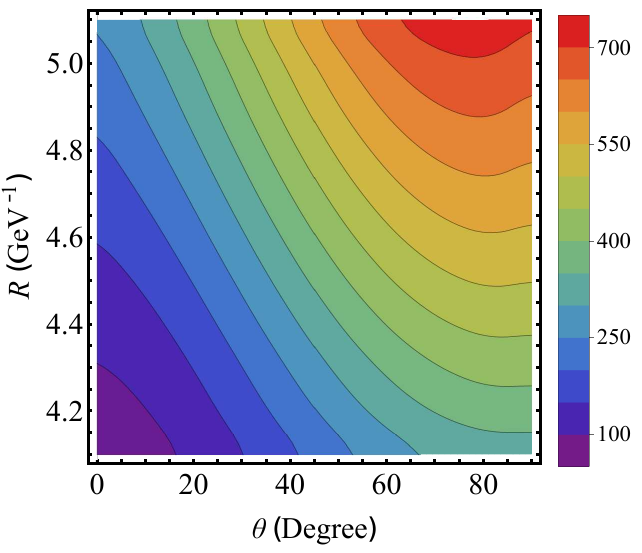}\\
	\end{tabular}
	\caption{(Color online.)
	{The contour plot of the calculated total decay widths of the $f_0(2020)$ as a mixing state of $s\bar{s}$ and $n\bar{n}$. $\theta$ is mixing angle.}}
	\label{contour}
\end{figure}

{With regard to the $f_0(2020)$, it is evident that its $\pi^0\pi^0$ decay mode, an experimental fact reported in Refs. \cite{GAMS:1998hws, Sarantsev:2021ein, Rodas:2021tyb, WA102:1999lqn, WA102:1997sum, Ropertz:2018stk}, should be considered. The $f_0(2020)$ cannot be treated solely as a pure $s\bar{s}$ state. While the calculated total decay width of the $f_0(2020)$ as a pure $s\bar{s}$ state aligns with the result from BESIII in Ref. \cite{BESIII:2022iwi}, it does not match the experimental width reported in Ref. \cite{BESIII:2022zel}, as shown in Fig. \ref{f02020} (a). Additionally, the partial decay widths are depicted in Figs. \ref{f02020} (b) and (c). Therefore, in our subsequent analysis, we incorporate the mixing of $s\bar{s}$ and $n\bar{n}$ to better understand the properties of the $f_0(2020)$. Here, {the $f_0(1812)$ is treated as the  orthogonal state of the $f_0(2020)$.} The mixing scheme is expressed as 
\renewcommand{\arraystretch}{1.5}
\begin{eqnarray}
	\left(
	\begin{array}{c}
		|f_0(2020)\rangle \\
		|f_0(1812)\rangle
	\end{array}
	\right)
	=\left(
	\begin{array}{ccc}
	\cos\theta & \sin\theta \\
	-\sin\theta	& \cos\theta
	\end{array}
	\right)
	\left(
	\begin{array}{c}
	|s\bar{s}\rangle  \\
	|n\bar{n}\rangle
	\end{array}
	\right).
\end{eqnarray}
Here, $\theta$ denotes the mixing angle. {The $f_0(1812)$, regarded as the second excitation of the $f_0(1370)$, was studied in our previous work alongside the $f_0(2100)$ \cite{Guo:2022xqu}. Therefore, in this discussion, we focus solely on the $f_0(2020)$.} In order to investigate the dependence of the OZI-allowed strong decay behavior of the $f_0(2020)$ on the mixing angle $\theta$ and $R$ value. It is demonstrated that this mixing effect contributes to an enhancement of the calculated total decay width of the $f_0(2020)$ (see Fig. \ref{contour} for more details), where the $R$ range is taken as $R=4.1-5.1$ GeV$^{-1}$ and the $\theta$ range is taken as $\theta=0^{\circ}-90^{\circ}$.

Our result shows that the total width of $f_0(2020)$ agrees with the result of $203\pm9$ MeV reported in Ref. \cite{BESIII:2022iwi} for $\theta$ ranges from $0^{\circ}$ to $40^{\circ}$ with variations in $R$. Conversely, the measurement of $436\pm9$ MeV by BESIII \cite{BESIII:2022zel} suggests that $\theta$ should range from $30^{\circ}$ to $90^{\circ}$, indicating a significant $n\bar{n}$ component. Assuming $R = 4.6$ GeV$^{-1}$ and considering $\theta = 15^{\circ}$ or $\theta = 50^{\circ}$, we derive different decay widths as reported by BESIII \cite{BESIII:2022zel,BESIII:2022iwi}, along with the corresponding partial decay widths of the primary decay channels, which are detailed in Table \ref{TAB3}.
Here, the $\pi\pi$, $\pi\pi(1300)$ and $\pi\pi(1800)$  channels are allowed in both results and have a considerable partial decay width, which can match the experimental observation \cite{GAMS:1998hws,Sarantsev:2021ein,Rodas:2021tyb,WA102:1999lqn,WA102:1997sum,GAMS:1998hws,Ropertz:2018stk}. There are also several main decay channels, including $\pi a_1(1260)$, $\eta\eta'$, $\eta\eta$, $\eta\eta(1295)$, $\eta'\eta'$, $KK$, $KK_1(1400)$ and $KK_1(1460)$.
However, there is still a lack of conclusive evidence for both results to examine which one of them is more appropriate. We hope for more precise experimental data on the resonance parameter of the $f_0(2020)$ and the decay behavior.}

\begin{table}[htbp]
	\centering
	\setlength{\tabcolsep}{8mm}
	\label{besiii}
	\caption{The typical partial decay widths of $f_0(2020)$ (in units of MeV) when taking typical $R = 4.6$ GeV $^{-1}$ and $\theta = 15^{\circ}, 50^{\circ}$. The channels with decay widths less than 5 MeV are not shown here, but their contributions are included in the total widths.}
	\label{TAB3}
	\begin{tabular}{ccc}
		\toprule[0.5pt]
			\toprule[0.5pt]
		Channels & $\theta = 15^{\circ}$ & $\theta = 50^{\circ}$\\
		\midrule[0.5pt]
		$\pi\pi$ & $1.44$ & $12.58$ \\
		$\pi\pi(1300)$ & $10.51$ & $92.09$   \\
	    $\pi\pi(1800)$ & $14.41$ & $126.25$  \\
	    $\pi a_1(1260)$ & $3.01$ & $26.40$  \\
	    $\eta\eta$ & $5.79$ & $9.56$  \\
	    $\eta\eta'$ & $5.86$ & $1.21$  \\
	    $\eta\eta(1295)$ & $23.50$ & $38.79$  \\
	    $\eta'\eta'$ & $12.43$ & $17.41$  \\
	    $KK$ & $20.38$ & $22.34$  \\
	    $KK_1(1400)$ & $17.50$ & $6.74$  \\
	    $KK_1(1460)$ & $82.75$ & $62.64$  \\
	    Total width & $203.92$  &  $429.40$ \\
		\bottomrule[0.5pt]
		\bottomrule[0.5pt]
	\end{tabular}
\end{table}

\begin{figure*}[htbp]
	\centering
	\begin{tabular}{c}
		\includegraphics[width=0.9\textwidth]{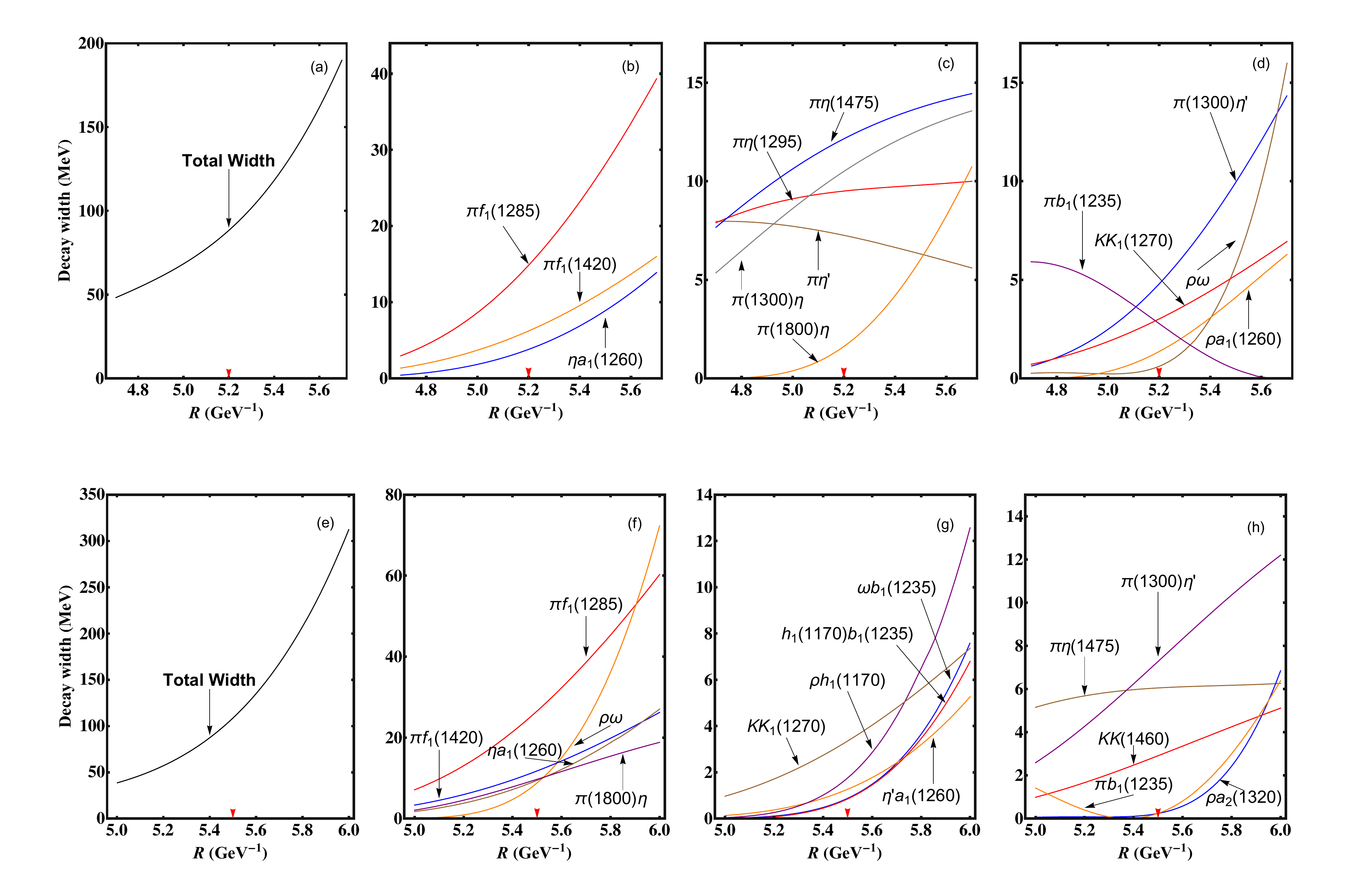}
	\end{tabular}
	\caption{(Color online.)
		The calculated partial widths and total decay widths of the {$a_0(2375)$} (a)-(d) and the predicted $a_0(2610)$ (e)-(i) dependent on $R$ values. The channels with decay widths less than 5 MeV are not shown here, but their contributions are included in the total widths. Here, the corresponding calculated $R$ values for these discussed scalar mesons are also marked with red arrows.  }
	\label{a02370}
\end{figure*}

In the following, we continue to discuss the strong decay behavior of the $f_0(2200)$. The $f_0(2200)$ is suitable to be categorized as the third radial excitation of the $f_0(1370)$, since the calculated total decay width of the $f_0(2200)$ is in agreement with experimental data, which can be reflected by the result obtained with $R=5.0-5.4$ GeV$^{-1}$ in Fig. \ref{f02020} (d). We continue to provide some valuable information about the main decay channels of the $f_0(2200)$ as shown in Fig. \ref{f02020} (e-f).
Here, its several main decay channels include $KK(1460)$, $KK_1(1270)$, and $KK_1(1400)$. 

The two-body OZI-allowed strong decay behavior of the $f_0(2470)$ is shown in Fig. \ref{f02470} (a)-(d). 
In the setting range of the $R$ value, our result of the total decay width is larger than the BESIII measurement \cite{BESIII:2022zel}. Since there is only one experimental result for the $f_0(2470)$, this discrepancy between our result and the BESIII data should be clarified in the future by checking the final states containing strange mesons.

The $f_0(2660)$, a predicted $f_0$ state, exhibits the decay behaviors detailed in Fig. \ref{f02470}, with the considered $R$ range extending from 5.0 GeV$^{-1}$ to 6.0 GeV$^{-1}$. As a broad state, the $f_0(2660)$ is calculated to have a total decay width of about 300 MeV, corresponding to the typical $R$ value of 5.5 GeV$^{-1}$. Its main decay modes include $K^*K^*$, $KK_1(1270)$, $KK_1(1460)$, $K^*K_1(1400)$, and $KK_1(1400)$. Experimental measurements could be used to further refine the theoretical calculations. Obviously, the search for the
predicted $f_0(2660)$ can be seen as a task for BESIII via the radiative decay of $J/\psi$.

\subsection{The predicted isovector $a_0(2375)$ and $a_0(2610)$}

The $a_0(2375)$ and $a_0(2610)$ are the two high-lying states in the $(n,M^2)$ trajectory of isovector scalar mesons that are still missing in experiments. In this work, we predict their two-body OZI-allowed strong decay behaviors by the QPC model, which are collected into Fig. \ref{a02370}.

The predicted mass of the $a_0(2375)$, assumed to be the fourth radial excitation of the $a_0(980)$, is estimated to be 2375 MeV by Eq. (\ref{nm}). As a high-lying state, the $a_0(2375)$ has abundant decay channels. The total decay width of the $a_0(2375)$, within the range $R=4.7-5.7$ GeV$^{-1}$, range from $50$ to $180$ MeV. Here, the $\pi f_1(1285)$, $\pi f_1(1420)$, and $\eta a_1(1260)$ channels have the major contribution to the total decay width. 
In Fig. \ref{a02370}, we also present the partial decay widths of these subordinate decay modes, the sum of which is significant to the total width of the $a_0(2375)$.


For the predicted $a_0(2610)$ state, its total and partial decay behavior are shown in in Fig. \ref{a02370}. Its total decay width is estimated to be $40-320$ MeV, corresponding to the $R$ range of $5.0\sim6.0 $ GeV$^{-1}$. Taking the typical 5.5 GeV$^{-1}$ as input, the total decay width is fixed at about 100 MeV.   
Our study can distinguish the main decay channels such as $\pi f_1(1285)$, $\pi f_1(1420)$, $\rho \omega$, $\eta a_1(1260)$ and $\pi(1800)\eta$ among these OZI-allowed strong decay channels, which should be preferentially considered when searching for the predicted $a_0(2610)$ state.


\section{Summary}

So far, our knowledge of the high-lying states of the scalar meson family is insufficient. By taking the opportunity of the observation of several scalar states from BESIII \cite{BESIII:2022zel,BESIII:2022iwi,BESIII:2022npc}, we perform the investigation of high-lying unflavored scalar mesons combined with a mass spectrum analysis by $(n,M^2)$ trajectory approach and the two-body strong decay calculation by the QPC model. Through this theoretical effort, the resonance parameters of these discussed scalar mesonic states can be obtained, which are crucial information for the research of hadron spectroscopy. 

In fact, the experimental effort should be made in the near future. A key point is the measurement of the resonance parameters of these scalar states. As shown in Fig.  \ref{comparison}, the results of the different experiments show significant differences, taking four scalar states $f_0(2020)$, $f_0(2330)$, $f_0(2100)$, and  $f_0(2200)$ as examples. With the accumulation of experimental data, we have reason to believe that this situation will be changed. 

Currently, we are entering the high-precision era of hadron spectroscopy with the upgrade of high-luminosity Large Hadron Collider, the running of Belle II, and the upcoming of upgrade plan of the Beijing Electron Positron Collider. These theoretical suggestions regarding high-lying scalar mesonic states presented in this work may be a new task for experimental study at BESIII.

\begin{acknowledgments}
This work is supported by  the National Natural Science Foundation of China under Grants No.~12335001 and No.~12247101, National Key Research and Development Program of China under Contract No.~2020YFA0406400, the 111 Project under Grant No.~B20063, Natural Science Foundation of Hebei Province under Grant No.~A2022203026, the Higher Education Science and Technology Program of Hebei Province under Contract No.~BJK2024176, the Research and Cultivation Project of Yanshan University under Contract No.~2023LGQN010, the fundamental Research Funds for the Central Universities, and the project for top-notch innovative talents of Gansu province.
\end{acknowledgments}

\bibliography{reference}

\end{document}